\documentclass[12pt,preprint]{aastex}      

\usepackage[latin1]{inputenc}
\usepackage{lscape}

\shorttitle{An optically thin view of HL Tau}
\shortauthors{Carrasco-Gonz\'alez et al.}

\begin{document}

\title{The VLA view of the HL Tau Disk - Disk Mass, Grain Evolution, and Early Planet Formation}

\author{Carlos Carrasco-Gonz\'alez\altaffilmark{1}, Thomas Henning\altaffilmark{2}, Claire J. Chandler\altaffilmark{3}, Hendrik Linz\altaffilmark{2}, Laura P\'erez\altaffilmark{3,4,5}, Luis F. Rodr\'{\i}guez\altaffilmark{1}, Roberto Galv\'an-Madrid\altaffilmark{1}, Guillem Anglada\altaffilmark{6}, Til Birnstiel\altaffilmark{2}, Roy van Boekel\altaffilmark{2}, Mario Flock\altaffilmark{7}, Hubert Klahr\altaffilmark{2}, Enrique Macias\altaffilmark{6},
Karl Menten\altaffilmark{4}, Mayra Osorio\altaffilmark{6}, Leonardo Testi\altaffilmark{8,9,10}, Jos\'e M. Torrelles\altaffilmark{11,12}, Zhaohuan Zhu\altaffilmark{13}}

\altaffiltext{1}{Instituto de Radioastronom\'{\i}a y Astrof\'{\i}sica UNAM,
Apartado Postal 3-72 (Xangari), 58089 Morelia, Michoac\'an, M\'exico. [c.carrasco,l.rodriguez,r.galvan]@crya.unam.mx}
\altaffiltext{2}{Max-Planck-Institut f\"ur Astronomie Heidelberg, K\"onigstuhl 17, 69117 Heidelberg, Germany. [henning,linz,birnstiel,boekel,klahr]@mpia.de}
\altaffiltext{3}{National Radio Astronomy Observatory, P.O. Box O, 1003 Lopezville Road, Socorro, NM 87801-0387, USA. cchandle@nrao.edu}
\altaffiltext{4}{Jansky Fellow of the National Radio Astronomy Observatory}
\altaffiltext{5}{Max-Planck-Institut f\"ur Radioastronomie Bonn, Auf dem H\"ugel 69, 53121 Bonn, Germany. [lperez,kmenten]@mpifr.de}
\altaffiltext{6}{Instituto de Astrof\'{\i}sica de Andaluc\'{\i}a (CSIC), Apartado 3004, 18080 Granada, Spain. [guillem,emacias,osorio]@iaa.es}
\altaffiltext{7}{Jet Propulsion Laboratory, California Institute of Technology,
4800 Oak Grove Drive, Pasadena, CA, 91109, USA. mario.flock@jpl.nasa.gov}
\altaffiltext{8}{European Southern Observatory, Karl-Schwarzschild-Str.~2, 85748 Garching bei M\"unchen, Germany. ltesti@eso.org}
\altaffiltext{9}{INAF-Osservatorio Astrofisico di Arcetri, Largo E. Fermi 5, I-50125 Firenze, Italy.} 
\altaffiltext{10}{Excellence Cluster ``Universe'', Boltzmann-Str.~2, 85748 Garching bei M\"unchen, Germany.}
\altaffiltext{11}{Institut de Ci\`encies de l'Espai (CSIC-IEEC) and Institut de Ci\`encies del Cosmos (UB-IEEC), Mart\'{\i} i Franqu\`es 1, 08028 Barcelona, Spain }
\altaffiltext{12}{The ICC (UB) is a CSIC-Associated Unit through the ICE}
\altaffiltext{13}{Department of Astrophysical Sciences, Princeton University, Princeton, NJ 08544, USA}

\begin{abstract}
 The first long-baseline ALMA campaign resolved the disk around the young star HL Tau into a number of axisymmetric bright and dark rings. Despite the very young age of HL Tau these structures have been interpreted as signatures for the presence of (proto)planets. The ALMA images triggered numerous theoretical studies based on disk-planet interactions, magnetically driven disk structures, and grain evolution. Of special interest are the inner parts of disks, where terrestrial planets are expected to form. However, the emission from these regions in HL~Tau turned out to be optically thick at all ALMA wavelengths, preventing the derivation of surface density profiles and grain size distributions. Here, we present the most sensitive images of HL~Tau obtained to date with the Karl G. Jansky Very Large Array at 7.0 mm wavelength with a spatial resolution comparable to the ALMA images. At this long wavelength the dust emission from HL Tau is optically thin, allowing a comprehensive study of the inner disk. We obtain a total disk dust mass of (1-3)$\times$10$^{-3}$ M$_\sun$, depending on the assumed opacity and disk temperature. Our optically thin data also indicate fast grain growth, fragmentation, and formation of dense clumps in the inner densest parts of the disk. Our results suggest that the HL~Tau disk may be actually in a very early stage of planetary formation, with planets not already formed in the gaps but in the process of future formation in the bright rings.
\end{abstract}

\section{Introduction}

 HL~Tau is a very young solar-type star surrounded by a dusty circumstellar disk and a remnant envelope. The object is located at a distance of $\sim$140~pc (Loinard et al. 2007), within the Taurus star-forming region. Showing all ingredients of a young system in the earliest stages of planet formation, HL~Tau has attracted a lot of attention over the years. For a summary of the early observational data and the results of the first comprehensive radiative transfer modeling we refer to D'Alessio et al. (1997) and Men'shchikov, Henning \& Fischer (1999).

 HL~Tau drives an ionized jet indicating on-going accretion (e.g., Pyo et al.~2005, Anglada et al. 2007). Early interferometric observations revealed that emission at cm wavelengths traces the radio counterpart of this collimated jet, while the emission at wavelengths $\lesssim$ 1.3 cm predominantly traces dust emission from a disk (Rodr\'{\i}guez et al. 1994, Wilner et al. 1996). This source attracted renewed interest after high angular resolution interferometric observations indicated that the HL Tau disk, despite its youth, may already be forming planets. Observations performed with the Combined Array for Research in Millimeter-wave Astronomy (CARMA) at 1.3 and 2.7~mm ($\sim$20-120~au resolution) suggested a gravitationally unstable disk which might undergo fragmentation (Kwon et al. 2011). Very Large Array (VLA) observations at 1.3 cm ($\sim$12~au resolution) revealed a compact structure in the disk at 65~au radius, interpreted as a protoplanet candidate (Greaves et al. 2008). Subsequent high sensitivity VLA observations at 7.0~mm ($\sim$7~au resolution) could not confirm this putative proto-planet, but found evidence for a depression at radius $\sim 10$~au in the radial density profile of the disk, which was interpreted as being related to the presence of an orbiting protoplanet (Carrasco-Gonz\'alez et al. 2009).

 With the long baselines of the Atacama Large Millimeter/submillimeter Array (ALMA) becoming available (ALMA Partnership et al. 2015a), this facility produced iconic images of the dust emission at 2.9, 1.3, and 0.87~mm from the HL~Tau disk ($\sim$3.5 to 10~au resolution), showing a number of axisymmetric bright and dark rings, most probably corresponding to high and low density concentric dust structures in the disk (ALMA Partnership et al. 2015b). The images immediately triggered numerous theoretical works in order to explain these remarkable structures. Planet-related explanations range from the presence of embedded sub-Jupiter mass planets (Picogna \& Kley 2015, Dipierro et al. 2015, Dong et al. 2015) to individual more massive planets (Gonzalez et al. 2015). Alternative explanations include magnetized disks without planets (Flock et al. 2015), fast pebble growth near condensation fronts (Zhang et al. 2015), and sintering-induced dust rings (Okuzumi et al. 2016). 

 The presence of massive planets ($\sim$10-15 M$_{\rm J}$) in two prominent dips in the dust distribution at $\sim$70~au was excluded utilizing deep direct L' band imaging with the Large Binocular Telescope (LBT) (Testi et al. 2015), but the presence of lower-mass planets in the disk is not yet excluded and remains an interesting possibility.

 Detailed radiative transfer analysis of the ALMA data shows that the emission from the various bright rings is probably optically thick, even at the longest ALMA wavelength of 2.9 mm (Pinte et al. 2016, Jin et al. 2016). The challenge of deriving density profiles and grain size distributions can only be circumvented by observations at even longer wavelengths where the disk will be optically thinner. In this Letter, we present new high-sensitivity Karl G.~Jansky VLA observations at 7.0~mm of the HL~Tau disk. This data provide a deeper view of the HL Tau disk, with an angular resolution comparable to the ALMA images.

\section{Observations and image analysis}

 We observed HL~Tau with the VLA of the National Radio Astronomy Observatory (NRAO)\footnote{The NRAO is a facility of the National Science Foundation operated under  cooperative agreement by Associated Universities, Inc.} using the Q band receivers in the C, B, and A configurations (see Table \ref{Tab1} for details). We observed the frequency range 39-47 GHz (central wavelength $\simeq$ 7.0~mm). Calibration of the data was performed with the data reduction package Common Astronomy Software Applications (CASA; version 4.4.0), using a modified version of the NRAO calibration pipeline. 

 Images were made with the CASA task {\tt CLEAN} using multi-scale multi-frequency synthesis that fits the emission with a Taylor series with nterms=2 during the deconvolution (Rau \& Cornwell 2011). Since our multi-configuration observations are sensitive to emission at very different scales (from $\sim$16$\arcsec\simeq$2240~au to 0$\farcs$05$\simeq$7~au), we made images with different angular resolutions by adjusting the Briggs {\tt robust} parameter (Briggs 1995) and the Gaussian uv-taper in {\tt CLEAN}. We also made images by splitting the 8 GHz band in two sub-bands of 4 GHz each (central wavelengths 6.7 and 7.3 mm). For comparison, images were aligned by assigning to the position of the central peak of the ALMA images the same absolute coordinates than in the VLA images, $\alpha$(J2000)=04$^h$31$^m$38$^s$426, $\delta$(J2000)=18$^\circ$13$\arcmin$57$\farcs$23. In Figures \ref{Fig1} and \ref{Fig2} we present the VLA images with different angular resolutions and comparisons with the ALMA images. 
 
 We obtained radial profiles of the intensity of the ALMA images and our most sensitive VLA image at 7.0~mm (natural weighting; beam size$\simeq$0$\farcs$067). For a proper comparison, we convolved all images to a common circular beam size of 0$\farcs$07 (the smallest beam size obtained from the public 2.9~mm ALMA uv-data with a uniform weighting). From these images, we obtained the average intensity within concentric elliptical rings of 0$\farcs$01 width at different radii, using the task {\tt IRING} of the Astronomical Image Processing System (AIPS). The dimension and orientation of the elliptical rings match those of the HL~Tau disk derived from the ALMA images (inclination angle, $i\simeq$46.72$^\circ$, and position angle, P.A.$\simeq$138.02$^\circ$; ALMA Partnership et al. 2015b). 
  
 Some contamination from free-free emission from the HL~Tau jet is expected at 7.0~mm. From our 6.7 and 7.3 mm sub-band images we obtain a spectral index of $\sim$1 at the center of the disk, consistent with comparable contributions from free-free and dust emission. To correct for the free-free contamination, we calibrated VLA A configuration archival data at 6 and 2 cm (project code: 12B-272) where emission is dominated by a partially optically thick radio jet in the NE-SW direction. From these images, the frequency dependence of the jet's flux density and angular size (major axis) can be expressed as $S_\nu \simeq 280 \times [\nu / 10.5~\rm{GHz}]^{0.4}$~$\mu$Jy and $\theta_{maj} \simeq 0\farcs10 \times [\nu / 10.5 ~ \rm{GHz}]^{-1}$, respectively, consistent with previous 7~mm observations (e.g. Wilner \& Lay 2000). Thus, an upper limit to the free-free contribution at 7.0~mm is obtained by extrapolating the flux density from cm wavelengths with a spectral index of 0.4, while a lower limit can be obtained by assuming free-free emission becomes optically thin at 2 cm (i.e., spectral index of $-$0.1 from 2 cm to shorter wavelengths). Therefore, we expect unresolved ($<$0$\farcs$07) free-free emission with a flux density in the range $\sim$200-400~$\mu$Jy corresponding to $\sim$35-65\% of the 7.0 mm emission at the disk center. This correction implies a larger uncertainty in the dust intensity at the center of the disk. 
 
 From the corrected intensity profiles, we derived brightness temperatures at each wavelength using the Planck equation for blackbody radiation, and we computed spectral indices between different wavelengths. We also derived profiles of optical depth and column density (assuming dust temperature power-laws and opacity, see \S3.1). Radial profiles are shown in Figure 3.
  
\section{Results and discussion}

\begin{figure}
\begin{center}
\includegraphics[width=\textwidth]{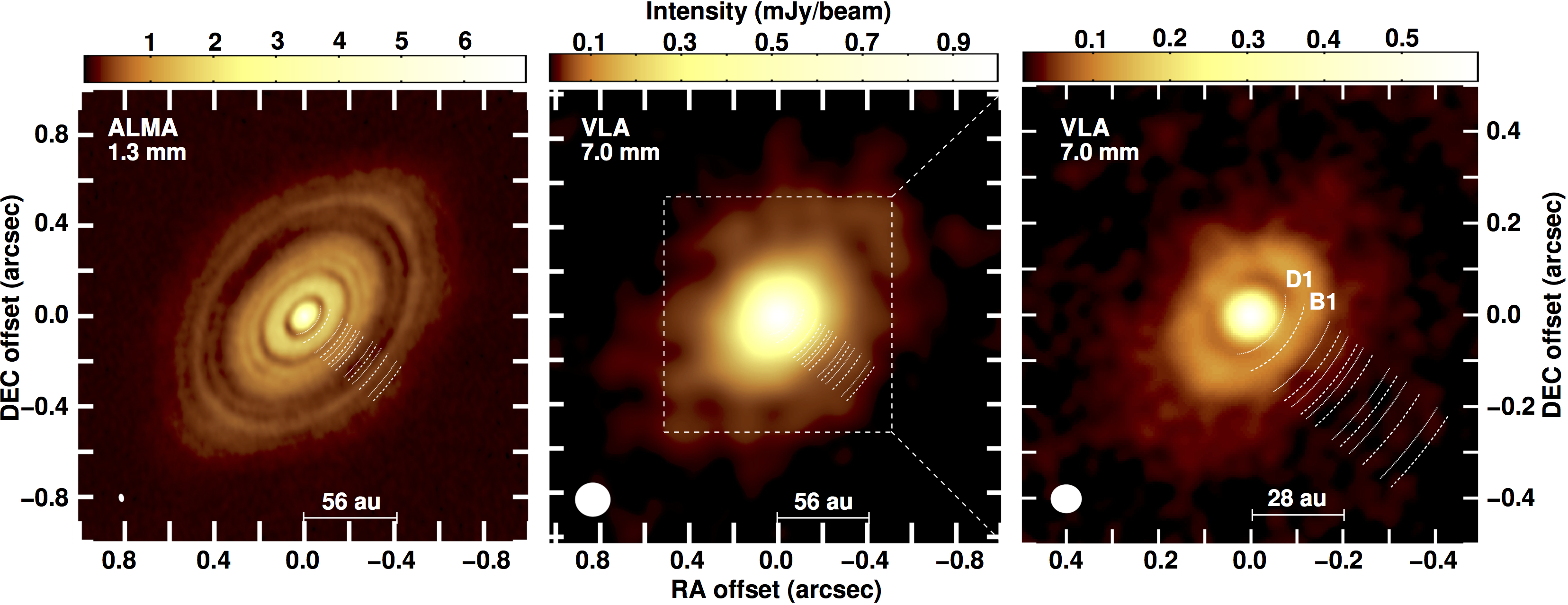}
\caption{\footnotesize{Comparison between the ALMA and VLA observations of the HL Tau disk. \textbf{Left:} ALMA image at 1.3~mm. \textbf{Center:} VLA image at 7.0~mm with an angular resolution of $\sim$20~au (0$\farcs$15; tapered image). \textbf{Right:} Close-up to the center of the disk. VLA image at 7.0~mm with an angular resolution of $\sim$10~au ($\sim$0$\farcs$07; natural weighting). In all panels, the positions of the reported dark (D1-D7; dotted lines) and bright rings (B1-B7; dashed lines) from the ALMA images (ALMA Partnership et al. 2015b) are shown. The inner disk and the first pair of dark (D1) and bright (B1) rings are clearly seen in the 7.0~mm images.}}
\label{Fig1}

\end{center}
\end{figure}

 The recent ALMA images of the HL Tau disk revealed several dark and bright rings (named D1-D7 and B1-B7, respectively) (Fig. \ref{Fig1}). Our new 7.0~mm VLA observations are the most sensitive and highest angular resolution observations of the HL~Tau disk performed to date at such a long wavelength. The low-angular resolution 7.0 mm image shows an elliptical source with similar size and orientation as the ALMA images (Fig. \ref{Fig1}). At higher angular resolution, the VLA is able to image with high signal-to-noise (S/N) ratio ($>$4-$\sigma$) the 7.0~mm emission from the inner half of the disk ($\lesssim$50~au; see Figs. \ref{Fig1} and \ref{Fig2}). In our 7.0~mm images we clearly identify several of the features seen in the ALMA images: the central disk and the first pair of dark (D1) and bright (B1) rings (Figs. \ref{Fig1}-\ref{Fig3}). 
 
 The importance of our sensitive 7.0~mm images is that, at such long wavelength, the emission has lower optical depth than in the ALMA data. This is especially critical for the study of the innermost part of the disk, where dust becomes opaque at all ALMA wavelengths and, as a consequence, physical properties are poorly constrained even with detailed modeling (e.g. Pinte et al. 2016, Jin et al. 2016). At the positions of the most opaque regions, the center of the disk and the first bright ring, the 7.0~mm brightness temperatures ($\sim$45 and 15 K, resp.) are $\sim$4 times lower than those of the ALMA 0.87~mm image ($\sim$130 and 60 K, resp.). Therefore, assuming that at 0.87~mm the emission from these two structures is optically thick, we obtain optical depths $\lesssim$0.4 at 7.0~mm. This imply that dust emission at 7.0~mm is well optically thin at all radii, even in the densest parts of the disk. 
 
 Our new high sensitivity 7.0 mm images of the HL Tau disk are an excellent basis for future comprehensive radiative transfer modeling to accurately obtain the physical properties of the disk. They are especially necessary in order to better constrain properties in the inner disk regions, where terrestial planets are thought to form, in principle. 
 
 In the following, we analyze our VLA images of the HL Tau disk to obtain direct rough estimates of the different physical parameters (e.g., mass and grain size distributions). We also analyze possible substructure in the disk and discuss our results in the context of planet formation. 

\subsection{Mass distribution}

 An accurate determination of the mass distribution in the HL~Tau disk requires detailed radiative transfer modeling. For this paper, we obtain first estimates by assuming a simple power-law for the dust temperature in the form $T_{dust}=T_{0} (R/R_{0})^{-q}$. While the exponent seems to be well constrained in the range $q$=0.5-0.6 by previous studies, there is large uncertainty in the reference temperature, with different proposed values in the range $T_0\simeq$70-140~K at $R_0$=10~au (e.g., Men'shchikov et al. 1999, Kwon et al. 2011, Pinte et al. 2016). For the dust opacity at 7.0~mm, we use a range of typical values for the disk-averaged opacity, $\kappa_{7mm}=$0.13-0.2 cm$^2$~g$^{-1}$ (e.g., Men'shchikov et al. 1999, P\'erez et al. 2012). Thus, at each radius, we calculate ranges for the optical depth and the dust column density taking into account these uncertainties (see Fig. \ref{Fig3}). Our calculations are consistent with the inner features of the disk being optically thick at all ALMA wavelengths, while at 7.0~mm the emission is optically thin at all radii (see Fig. \ref{Fig3}b). We estimated values of the dust column density around $\sim$1~g~cm$^{-2}$ at the center of the disk (see Fig. \ref{Fig3}c). This suggest a denser disk at inner radii ($<$50~au) than previously obtained by detailed modeling (e.g., Pinte et al. 2016 predict $\lesssim$0.2~g~cm$^{-2}$ at the center of the disk). 

 We also estimated dust masses for the inner disk (ID) and the bright rings (B1 to B6; see Table \ref{Tab2}). For those features which are optically thin in the ALMA images, i.e. B2-B6, we obtain dust masses consistent with previous estimations (Pinte et al. 2016). However, our optically thin 7.0~mm data suggest large dust masses for the inner disk and the first bright ring (B1) for which only lower limits were obtained previously (see Table \ref{Tab2}). Finally, we estimate that the total dust mass of the disk is within the range (1-3)$\times$10$^{-3}$~M$_\sun$, which is also somewhat larger than previous estimates, $\sim$(0.3-1)$\times$10$^{-3}$~M$_\sun$ (e.g.,  Men'shchikov et al. 1999, D'Alessio et al. 1997, Kwon et al. 2008, Pinte et al. 2016). 
 
\subsection{Dust particle-size distribution}

 Grain growth and mixing lead to changes in particle-size distribution and dust composition throughout the disk (Henning \& Meeus 2011). This has been recently studied in several objects for which segregation by particle-size (e.g. Menu et al. 2014) and radial changes in dust optical properties (e.g. Guilloteau et al. 2011, P\'erez et al. 2012, 2015) are observed. 
 
 The fully resolved ALMA and VLA images of the HL~Tau disk offer now an excellent opportunity for a detailed study of the properties of the particle-size distribution in a very young disk. In particular, changes in the dust properties can be inferred from changes in the spectral index of the emission, $\alpha$, but only for optically thin emission in the Rayleigh-Jeans regime (e.g. Beckwith et al. 2000). When derived from the short ALMA wavelengths, the observed radial variations of $\alpha$, from $\sim$2 to 2.5 (Figure \ref{Fig3}d), reflect high optical depths inwards of $\sim$50~au. Thus, these ALMA observations cannot be used to infer grain growth in the densest, inner disk regions. In contrast, the observed radial variations of $\alpha$ derived from the two most optically thin wavelengths, 7.0 and 2.9 mm, show a different behavior: (1) at all radii, except at the location of the dark gap D5, we obtain $\alpha_{7.0-2.9mm}>\alpha_{1.3-0.87mm}$, consistent with the emission at shorter wavelengths being more optically thick and not in the R-J regime, and (2) a clear gradient in $\alpha_{7.0-2.9mm}$ is observed between $\sim$10-50 au, consistent with a change in the dust optical properties and a differential grain-size distribution, with larger grains at smaller radii. Similar results have been inferred for a number of more evolved disks (e.g. P\'erez et al. 2012, 2015; Tazzari et al. 2016, Menu et al. 2014).

\subsection{Substructure in the first bright ring}
 
\begin{figure}
\begin{center}
\includegraphics[width=\textwidth]{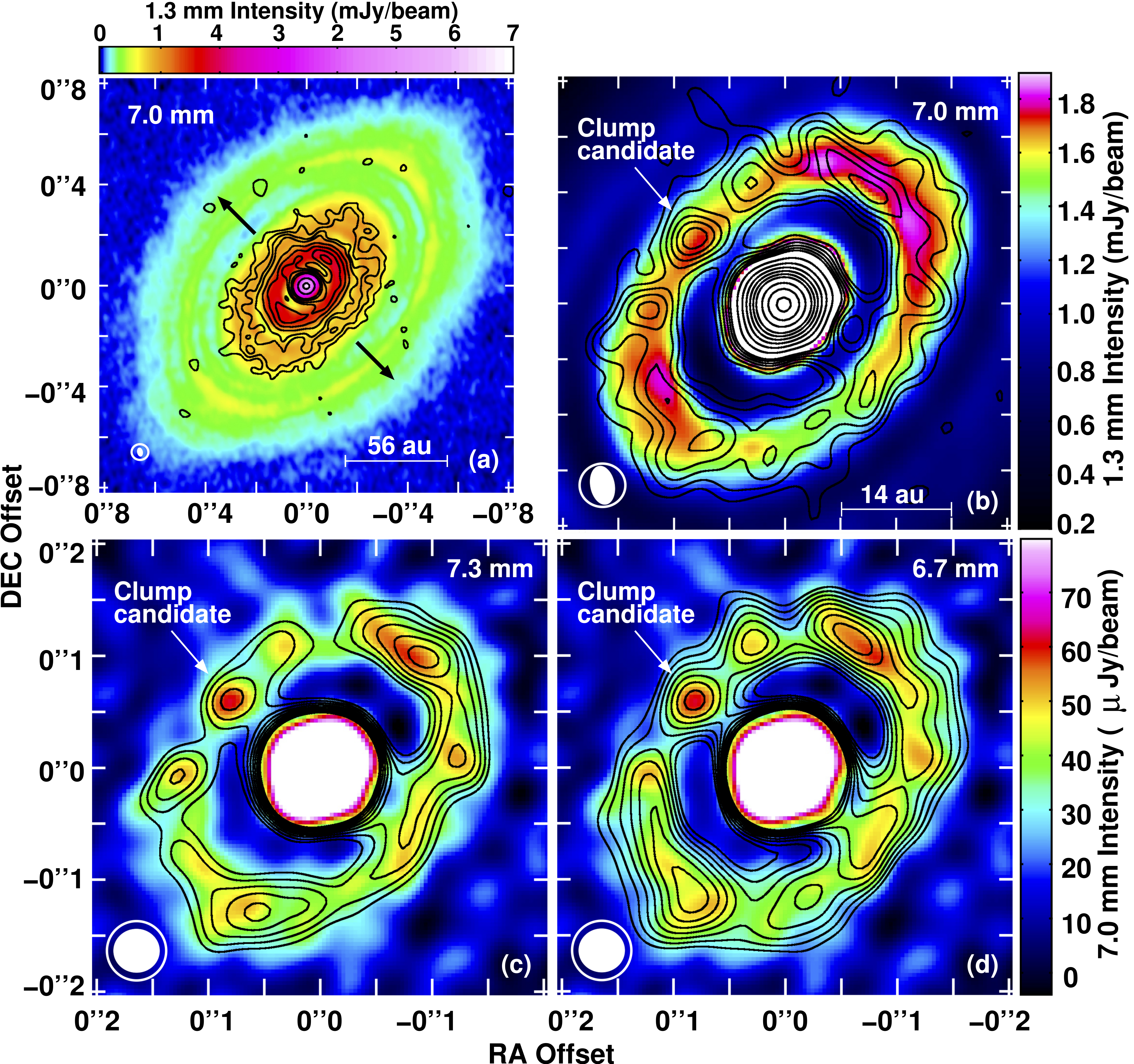}
\caption{\footnotesize{\textbf{(a)} Superposition of the VLA 7.0 mm image (contours; naturally weighted image; beam size$\simeq$0$\farcs$067) over the ALMA 1.3~mm image (color scale). Contour levels are $-$4, 4, 6, 8, 10, 15, 20, 25, 30, 40, 50, 100, and 150 times the rms of the 7.0~mm map, 3.5 $\mu$Jy~beam$^{-1}$. The two arrows mark the direction of the collimated jet at a P.A. of $\sim$45$^\circ$ (Anglada et al. 2007). \textbf{(b)} A close-up to the center of the disk. Color scale is the ALMA 1.3~mm image and contours are from the high angular resolution VLA 7.0~mm image (robust 0 weighted image; beam size$\simeq$0$\farcs$04). Contour levels are 3, 4, 5, 6, 7, 8, 12, 16, 20, 24, 28, 32, 40, 48, and 56 times 7~$\mu$Jy~beam$^{-1}$. \textbf{(c) and (d):} Comparison between sub-bands contour images at 7.3~mm and 6.7~mm (robust 0.4 weighting; beam sizes$\simeq$0$\farcs$053) over the 7.0~mm color scale image. Contours in both panels are 8, 9, 10, 11, 12, 13, 14, 15, and 16 times 6~$\mu$Jy~beam$^{-1}$.}}
\label{Fig2}

\end{center}
\end{figure}

\begin{figure}
\begin{center}
\includegraphics[width=\textwidth]{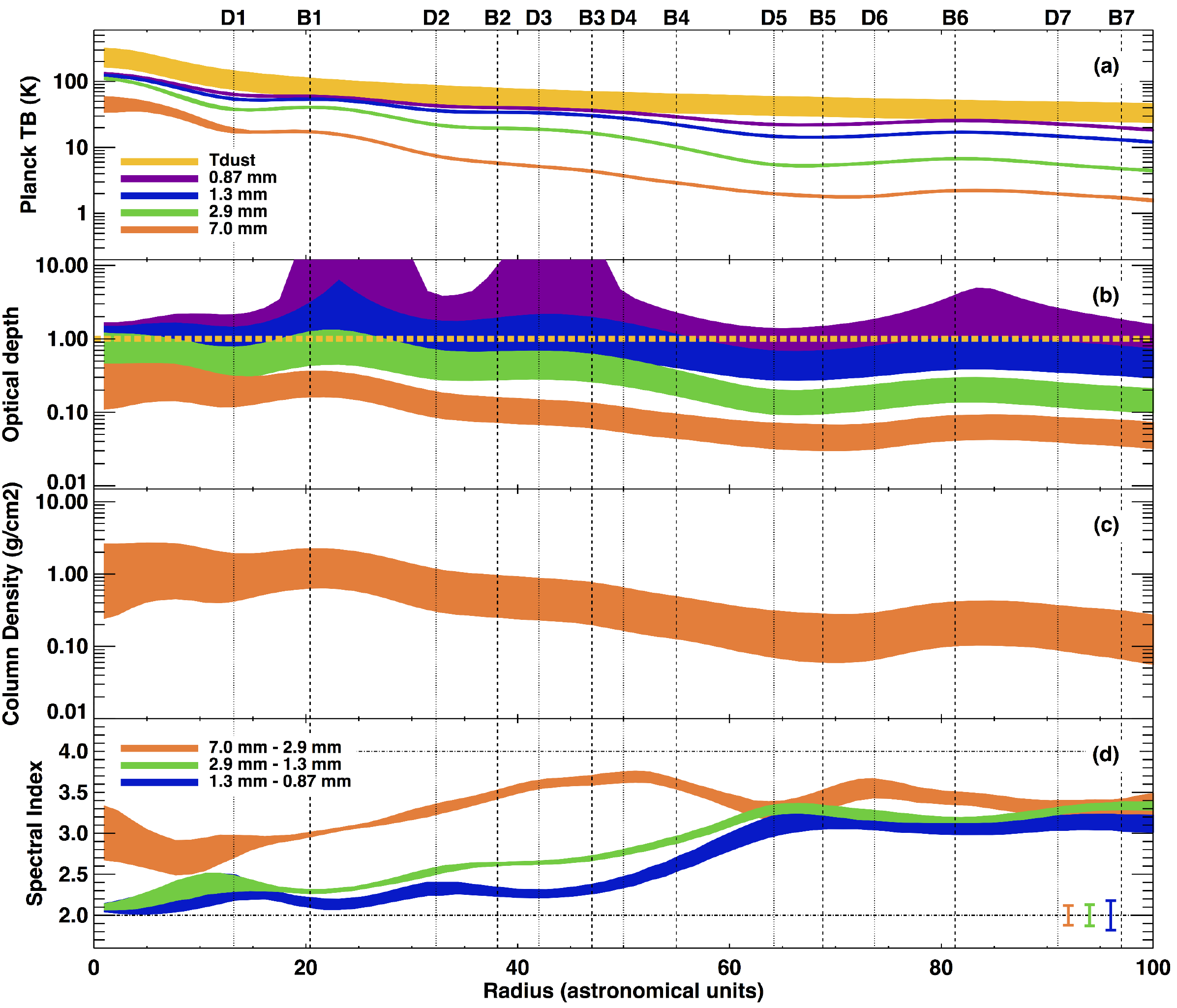}
\caption{\footnotesize{Radial profiles of several quantities in the HL Tau disk. In all panels, the width of the lines represents the 1-$\sigma$ uncertainty of each quantity. \textbf{(a)} Brightness temperature at different wavelengths (2.9, 1.3 and 0.87~mm from ALMA; 7.0~mm from VLA). Obtained by averaging the intensity in concentric ellipses over the ALMA and VLA images convolved to a common circular beam size of 0$\farcs$07. A dust temperature power-law, also convolved to a beam size of 0$\farcs$07, is also shown (see \S3.1). \textbf{(b)} Optical depth obtained by assuming the dust temperature profile in panel (a). The thick dashed horizontal line marks the threshold between optically thin ($<$1) and optically thick ($>$1) emission. \textbf{(c)} Column density profile obtained from the 7.0~mm data and the dust temperature profile. \textbf{(d)} Spectral index profiles between several wavelengths. Error bars in bottom-right corner indicate uncertainties due to absolute flux calibration (not affecting spectral index gradient). }}
\label{Fig3}

\end{center}
\end{figure}

 Some more evolved transitional disks, when observed at long wavelengths, show knotty rings of dust emission (e.g., HD~169142; see Fig. 1c in Osorio et al. 2014). Our highest angular resolution VLA images of the young HL~Tau disk also reveal an interesting knotty and not axisymmetric morphology of the first bright ring (B1) (see Fig. \ref{Fig2}b). Most of the knots seem to be consistent with small (1-$\sigma$) fluctuations of the brightness (due to the rms noise of the map), suggesting a structure with a roughly uniform brightness. However, to the NE, there is a compact knot (labeled as ``clump candidate" in Figs. \ref{Fig2}b-d) whose morphology is very different to the rest of the ring. We made several images with different angular resolutions and different bandwidths (see \S2), and noted that, while the knotty emission changes significantly in different images (consistent with being small fluctuations in a uniform structure), the NE knot is clearly identifiable and the most compact knot in all the images (see Figs. \ref{Fig2}c and \ref{Fig2}d). This knot also coincides with a local intensity maximum in the 1.3 mm ALMA image (see Fig. 2b). All this suggests that the 7.0~mm NE knot traces a real structure in the first bright ring. 
 
 We note that the position of the NE knot is coincident with the direction of the jet, and thus it could be related to a local increase in the flux density because of the foreground free-free emission. However, at 7.0~mm, the free-free emission from the jet is expected to be confined within the inner disk (see \S2). Moreover, from our sub-band images we derive a spectral index $\alpha_{6.7-7.3 mm}$=2.5$\pm$0.4, suggesting that the emission from this knot is dominated by thermal dust emission. We also noted that this spectral index is slightly smaller than the average spectral index obtained for B1 from 7.0 and 2.9~mm (see Fig. \ref{Fig3}), suggesting an accumulation of larger dust grains in the clump. It is known that dense rings could undergo vortex formation by the Rossby Wave Instability and efficiently concentrate large particles (Meheut 2012), then, we speculate that the NE knot traces a dense dust clump formed within the massive bright ring. In this case, we estimate a dust mass in the range 3-8~M$_\earth$ for this clump.  
 
\subsection{Planet formation in the HL Tau disk}
 
 The presence of dark and bright concentric rings has been commonly interpreted as the result of planet formation already ongoing in the HL Tau disk. However, since HL~Tau is a very young T~Tauri star, the presence of several (proto)planets sufficiently massive to carve holes in the disk at this early stage is somewhat surprising. On the other hand, alternative formation mechanisms, not requiring the presence of protoplanets, seem also possible. Moreover, sensitive searches for massive (proto)planets in the outer dark rings have yielded negative results (see \S1 and references therein). 
 
 We propose a scenario in which the HL~Tau disk may have not formed planets yet, but rather is in an initial stage of planet formation. Instead of being caused by (proto)planets, the dense rings could have been formed by an alternative mechanism. Our 7.0~mm data suggest that the inner rings are very dense and massive, and then, they can be gravitationally unstable and fragment. It is then possible that the formation of these rings result in the formation of dense clumps within them like the one possibly detected in our 7.0~mm image. These clumps are very likely to grow in mass by accreting from their surroundings, and then they possibly represent the earliest stages of protoplanets. In this scenario, the concentric holes observed by ALMA and VLA would not be interpreted as a consequence of the presence of massive (proto)planets. Instead, planets may be just starting to form in the bright dense rings of the HL~Tau disk. 

\noindent \emph{Acknowledgments.} CC-G, LFR and RG-M acknowledge support from UNAM-DGAPA PAPIIT IA101715 and IA102816. LMP acknowledges support from the Alexander van Humboldt Foundation. TB acknowledges support from the DFG grant (KL 1469/13-1). GA, EM, MO, and JMT acknowledge support from MINECO and FEDER funds. This paper makes use of the following ALMA data: ADS/JAO.ALMA\#2011.0.00015.SV. ALMA is a partnership of ESO (representing its member states), NSF (USA) and NINS (Japan), together with NRC (Canada), NSC and ASIAA (Taiwan), and KASI (Republic of Korea), in cooperation with the Republic of Chile. The Joint ALMA Observatory is operated by ESO, AUI/NRAO and NAOJ.

\begin{deluxetable}{lccc}
\tablewidth{0pt}
\tablecaption{VLA observations at 7 mm\label{Tab1}}
\startdata
\hline \hline 
Obs.        &  Project &        & On-source  \\
Date        &   Code   & Conf.  & total time \\ \hline
2014-Dec-07 & 14B-485  &   C    &   1.7 h    \\
2015-Feb-15 & 14B-485  &   B    &   1.6 h    \\
2015-Aug-13 & 14B-487  &   A    &   1.1 h    \\
2015-Aug-25 & 14B-487  &   A    &   1.1 h    \\
2015-Sep-19 & 14B-487  &   A    &   1.7 h    \\
2015-Sep-20 & 14B-487  &   A    &   3.8 h    \\
2015-Sep-21 & 14B-487  &   A    &   3.4 h    \\
\enddata

\tablecomments{The bandpass, flux density and phase calibrators were 3C84, 3C147 and J0431+1731, respectively. The phase calibrator was observed every 3 minutes for the C configuration observations, and every 2 minutes for the B and A configurations observations. We estimate a 10\% uncertainty for the absolute flux calibration. Quoted flux errors in discussion are statistical uncertainties. Self-calibration was achieved. The phase center was $\alpha$(J2000)=04$^h$31$^m$38$^s$429, $\delta$(J2000)=18$^\circ$13$\arcmin$57$\farcs$29. }

\end{deluxetable}

\begin{deluxetable}{ccccc}
\tablewidth{0pt}
\tablecaption{Parameters of the disk features\label{Tab2}}
\startdata
\hline \hline 
             &          &  Flux Density &                      &                     \\
Covered Disk &  Radius  &   at 7.0 mm   & \multicolumn{2}{c}{Dust Mass (M$_\earth$)} \\
Feature      &   (au)   &    (mJy)      & This paper$^{\rm a}$ & Pinte et al.$^{\rm b}$\\ \hline
ID           & $<$13    & 0.61 $\pm$ 0.04 & 10 - 50     & $>$2.3 \\
B1           & 13 - 32  & 1.45 $\pm$ 0.02 & 70 - 210    & $>$47 \\
B2           & 32 - 42  & 0.48 $\pm$ 0.01 & 30 - 90     & 30 - 69\\
B3           & 42 - 50  & 0.35 $\pm$ 0.01 & 20 - 80     & 14 - 37 \\
B4           & 50 - 64  & 0.36 $\pm$ 0.01 & 30 - 90     & 40 - 82 \\
B5           & 64 - 74  & 0.18 $\pm$ 0.01 & 10 - 50     & 5.5 - 8.7 \\
B6           & 74 - 90  & 0.45 $\pm$ 0.01 & 40 - 140    & 84 - 129 \\
\enddata

\tablenotetext{a}{Calculated by integration of the column density profile obtained from the 7.0~mm data between adjacent dark rings.}
\tablenotetext{b}{Calculated by radiative transfer modeling of the ALMA images by Pinte et al. (2016).}

\end{deluxetable}

\end{document}